\documentclass[11pt]{article}

\newcommand{\const}{\,{\rm const}\,}

\def\be{\begin{equation}}
\def\ee{\end{equation}}
\def\bea{\begin{eqnarray}}
\def\eea{\end{eqnarray}}

\def\ell{l}

\title{\bf{Near-horizon symmetries of extremal black holes}}

\author{Hari K. Kunduri \\School of Physics and Astronomy, University of Nottingham, NG7 2RD, UK \\h.k.kunduri@nottingham.ac.uk \\ \\ James Lucietti \\ Centre for Particle Theory, Department of Mathematical Sciences, \\ University of Durham, South Road, Durham, DH1 3LE, UK\\ james.lucietti@durham.ac.uk \\  \\Harvey S. Reall \\ School of Physics and Astronomy, University of Nottingham, NG7 2RD, UK \\harvey.reall@nottingham.ac.uk \\ }
\date{29 May, 2007 ({\small DCPT-07/25})}

\begin{document}

\maketitle

\begin{abstract}
Recent work has demonstrated an attractor mechanism for extremal
rotating black holes subject to the assumption of a near-horizon
$SO(2,1)$ symmetry. We prove the existence of this symmetry for any
extremal black hole with the same number of rotational symmetries as
known four and five dimensional solutions (including black rings).
The result is valid for a general two-derivative theory of gravity
coupled to abelian vectors and uncharged scalars, allowing for a
non-trivial scalar potential. We prove that it remains valid in the presence of higher-derivative corrections. We show that $SO(2,1)$-symmetric near-horizon solutions can be analytically continued to give $SU(2)$-symmetric black hole
solutions.
For example, the near-horizon limit of an extremal 5D
Myers-Perry black hole is related by analytic continuation to a
non-extremal cohomogeneity-1 Myers-Perry solution.

\end{abstract}

\section{Introduction}

The ``attractor mechanism'' is the phenomenon that the entropy of an
extremal black hole cannot depend on any moduli of the theory. It
was initially discovered for supersymmetric black holes~\cite{FKS,Strominger,FK}, then realized that it still applies
in the presence of certain higher-derivative corrections~\cite{hd1,hd2,hd3}, and most recently
extended to non-supersymmetric black holes \cite{sen1, nonsusyatt, Kallosh}.
This has led to an explanation \cite{agm,dab} of the success of string
theory calculations of the entropy of certain non-supersymmetric
extremal black holes \cite{KLMS,HLM,Dabholkar,TT,EH,EM}.

Most studies of the attractor mechanism have concerned static,
spherically symmetric, black holes. However, there has been recent
interest in extending this to more general extremal black holes \cite{rotatt}.

Any extremal black hole admits a near-horizon limit \cite{hsr}. For known solutions, the isometry group of the black hole is enhanced in this limit. For example, the near-horizon geometry of the extremal Kerr black hole is~\cite{BH}
\be ds^2
=\frac{(1+\cos^2\theta)}{2}\left[ -\frac{r^2}{r_0^2}dv^2+2dvdr
+r_0^2 d\theta^2\right]+
\frac{2r_0^2\sin^2\theta}{1+\cos^2\theta} \left(
d{\phi}+\frac{r}{r_0^2}dv \right)^2 \ee
where $r_0>0$. The first two terms in square brackets are the metric of 2d anti-de Sitter space (AdS), which has isometry group $O(2,1)$. In fact (as we shall explain later), $O(2,1)$ extends to a symmetry of the full metric, so the full isometry group is $O(2,1) \times U(1)$ where the $U(1)$ arises from the axisymmetry of the black hole (generated by $\partial/\partial \phi$) \cite{BH}. Other 4d examples are considered in \cite{rotatt}, with the conclusion that they also have $O(2,1) \times U(1)$ isometry group in the near-horizon limit.

Similarly, the near-horizon geometry of the extremal 5d Myers-Perry \cite{MyersPerry} black hole has $O(2,1) \times U(1)^2$ isometry group \cite{BH}, where the $U(1)^2$ arises from the two rotational symmetries of this black hole. The original 2-parameter black ring solution \cite{ring} does not admit an extremal limit but its 3-parameter generalization \cite{Pomeransky} does, as does the dipole ring solution \cite{Emparan}. We shall see that the near-horizon geometries of these extremal solutions also have $O(2,1) \times U(1)^2$ symmetry.

In all of these examples, the $O(2,1)$ symmetry arises because the near-horizon geometry involves a fibration over $AdS_2$. For solutions with non-trivial Maxwell fields, the Maxwell field strengths are invariant only under the $SO(2,1)$ subgroup of $O(2,1)$ that preserves orientation in $AdS_2$. Hence the full near-horizon solution has symmetry group $SO(2,1) \times U(1)^{D-3}$ for $D=4,5$.

If one {\it assumes} the existence of this $SO(2,1)$ symmetry in general then one can extend the attractor mechanism beyond the static, spherically symmetric case to general extremal black holes \cite{rotatt}. At first sight, the assumption of $SO(2,1)$ symmetry appears rather strong since, as we shall explain, a general near-horizon geometry possesses only a 2d non-abelian isometry group. However, we shall show that $SO(2,1)$ emerges dynamically, as a consequence of the Einstein equation, subject to the assumption that the black hole in question admits the same number ($D-3$) of rotational symmetries as known black hole solutions in $D=4,5$ dimensions.

We shall work with a general 2-derivative theory describing Einstein gravity coupled to abelian vectors $A^I$ ($I=1\ldots N$) and uncharged scalars $\phi^A$ ($A=1 \ldots M$) in $D=4,5$ dimensions, with action
\be
\label{gentheory}
 S = \int d^D x \sqrt{-g} \left(R - \frac{1}{2} f_{AB}(\phi) \partial_\mu \phi^A  \partial^\mu \phi^B - V(\phi) - \frac{1}{4} g_{IJ}(\phi) F^I_{\mu\nu}  F^{J\mu\nu} \right) + S_{{\rm top}},
\ee
where $F^I \equiv dA^I$, $V(\phi)$ is an arbitrary scalar potential (which allows for a cosmological constant), and
\be\label{top4d}
 S_{{\rm top}} = \frac{1}{2} \int h_{IJ}(\phi) F^I \wedge F^J \qquad \mbox{\rm if $D=4$},
\ee
or
\be\label{top5d}
 S_{{\rm top}} = \frac{1}{6} \int C_{IJK}F^I \wedge F^J \wedge A^K \qquad \mbox{\rm if $D=5$},
\ee
where $C_{IJK}$ are constants.

This encompasses many theories of interest, e.g., vacuum gravity
with a cosmological constant, Einstein-Maxwell theory, and various
(possibly gauged) supergravity theories arising from
compactification from ten or eleven dimensions. Furthermore, we
shall not restrict attention only to asymptotically flat black
holes. For example, our results will apply equally to asymptotically
anti-de Sitter black holes. The first main result of this paper can be
summarised in the following:

\bigskip
\noindent {\bf Theorem 1.} Consider an asymptotically flat, or anti-de Sitter, extremal black hole solution of the above theory. Assume that it has $D-3$ rotational symmetries. Then the near-horizon limit of this solution has a global $G_3 \times U(1)^{D-3}$ symmetry, where $G_3$ is either $SO(2,1)$ or (the orientation-preserving subgroup of) the 2d Poincar\'e group. The Poincar\'e-symmetric case is excluded if $f_{AB}(\phi)$ and $g_{IJ}(\phi)$ are positive definite, the scalar potential is non-positive, and the horizon topology is not $T^{D-2}$.
\bigskip

\noindent {\it Remarks}:

\begin{enumerate}
\item  The asymptotic boundary conditions are only required at one point in the proof, where we use the fact that the generator of each rotational symmetry must vanish somewhere in the asymptotic region (on the ``axis" of the symmetry) to constrain the Maxwell fields. The theorem is true for any asymptotic boundary conditions with this property.\footnote{A $D=5$ theory with Kaluza-Klein boundary conditions could violate this condition (if one of the rotational Killing fields were tangent to the Kaluza-Klein circle at infinity then it would not vanish anywhere in the asymptotic region). However, in this case one could simply apply our theorem to the $D=4$ theory resulting from dimensional reduction.}
\item  The existence of a single rotational symmetry seems reasonable because of the ``stationary implies axisymmetric" theorem,\footnote{
That has been proved for $D=4$ Einstein gravity coupled to ``reasonable" matter (obeying the weak energy condition with hyperbolic equations of motion) and asymptotically flat boundary conditions \cite{hawkingellis}. It has recently been extended to $D>4$ and asymptotically anti-de Sitter boundary conditions~\cite{Wald}.}
although this has only been proved for {\it non-extremal} rotating black holes. There is no general argument for the existence of {\it two} rotational symmetries in $D=5$ but all known solutions have this property.

\item We will show that generic orbits of the symmetry group have the structure of $T^{D-3}$ fibred over over a 2d maximally symmetric space, i.e., $AdS_2$, $dS_2$ or $R^{1,1}$. $AdS_2$ and $dS_2$ give $SO(2,1)$ symmetry whereas $R^{1,1}$ gives Poincar\'e symmetry. We can exclude the $dS_2$ and $R^{1,1}$ cases subject to the additional assumptions mentioned, which ensure that the theory obeys the strong energy condition. This encompasses many theories of interest e.g. theories for which the scalars are all moduli, or various gauged supergravity theories. The assumption that the horizon topology is non-toroidal (which is not needed if the scalar potential is strictly negative) seems reasonable because of the black hole topology theorem, which has been proved for Einstein gravity with a variety of asymptotic boundary conditions and restrictions on the energy-momentum tensor.\footnote{In $D=4$ it has been proved for matter obeying the null energy condition and asymptotically flat \cite{chruscielwald} or asymptotically anti-de Sitter \cite{gallowaycensor} boundary conditions. In $D=5$ it has been proved for matter obeying the dominant energy condition and asymptotically flat boundary conditions \cite{Galloway}.}
\end{enumerate}

Much of the interest in the attractor mechanism stems from the fact
that it applies to any local, generally-covariant, theory, not just
second-order gravity \cite{sen1}. Therefore it is important to
examine how higher-derivative corrections affect our result. Our second main result is the following:

\bigskip

\noindent {\bf Theorem 2.} Consider a general theory of gravity coupled to abelian vectors $A^I$ and uncharged scalars $\phi^A$ with action
\be
\label{eqn:hdaction}
 S = S_2 + \sum_{m \ge 1} \lambda^m \int \sqrt{-g} {\cal L}_m,
\ee
where $S_2$ is the 2-derivative action above, $\lambda$ is a coupling constant, and ${\cal L}_m$ is constructed by contracting (derivatives of) the Riemann tensor, volume form, scalar fields and Maxwell fields in such a way that the action is diffeomorphism and gauge-invariant. Consider an extremal black hole solution of this theory obeying the same assumptions as in Theorem 1. Assume that there is a regular horizon when $\lambda=0$ with $SO(2,1) \times U(1)^{D-3}$ near-horizon symmetry (as guaranteed by Theorem 1), and that the near-horizon solution is analytic in $\lambda$. Then the near-horizon solution has $SO(2,1) \times U(1)^{D-3}$ symmetry to all orders in $\lambda$.

\bigskip

Hence our result is stable with respect to higher-derivative corrections. However, it does not apply to so-called ``small" black holes~\cite{SBH1,SBH2,SBH3,SBH4,SBH5}, for which existence of a horizon depends on the higher-derivative terms in the action, i.e., it requires $\lambda \ne 0$.

The above theorems are proved in section 2. Section 3 discusses some examples of 5-dimensional near-horizon geometries. In particular, we discuss the near-horizon geometries of extremal Myers-Perry black holes and black rings. The near-horizon geometry of an extremal vacuum black ring turns out the be the same as that of an extremal boosted Kerr black string. We shall see that the $SO(2,1) \times U(1)^2$-invariant near-horizon geometries of Myers-Perry and black ring solutions can be analytically continued to give stationary solutions with  $SU(2) \times U(1) \times R$ symmetry, where $R$ denotes time translations. For example, the near-horizon geometry of a cohomogeneity-2 Myers-Perry solution can be analytically continued to give a non-extremal cohomogeneity-1 (equal angular momenta) Myers-Perry solution. Similarly, the near-horizon geometry of an extremal dipole ring can be continued to give a Kaluza-Klein black hole. Finally, we determine the most general $SO(2,1) \times U(1)^{D-3}$-symmetric vacuum near-horizon geometry by exploiting the fact that the analytically continued version of this problem is to find the general stationary, spherically symmetric solution of Kaluza-Klein theory, which was solved in \cite{Maison2}.

\section{$SO(2,1)$ symmetry}

\subsection{Near-horizon limit}

The event horizon of a stationary, non-extremal, black hole must be a Killing horizon. We shall assume that this is also true for extremal black holes. In the neighbourhood of the horizon, one can introduce Gaussian null coordinates $(v,r,x^a)$, in which the metric takes the form (see e.g. \cite{racz})
\be
 ds^2 = r^2 F(r,x) dv^2 + 2 dv dr + 2 r h_a (r,x) dv dx^a + \gamma_{ab}(r,x) dx^a dx^b,
\ee
where $\partial/\partial v$ is Killing, the horizon is at $r=0$ and $x^a$ are coordinates on a $(D-2)$-dimensional spatial cross-section of the horizon. The functions $F$, $h_a$ and $\gamma_{ab}$ are continuous functions of $r$. Extremality implies that $g_{vv}$ is ${\cal O}(r^2)$.

The {\it near-horizon limit} is defined by \cite{hsr}
\be
\label{eqn:limit}
v \rightarrow v/\epsilon,  \qquad r \rightarrow \epsilon r
\ee
and $\epsilon \rightarrow 0$, after which we obtain the near-horizon geometry
\be
\label{eqn:nearhor}
 ds^2 = r^2 F(x) dv^2 + 2 dv dr + 2 r h_a (x) dv dx^a + \gamma_{ab}(x) dx^a dx^b,
\ee with $F(x) \equiv F(0,x)$ etc. The near-horizon geometry is
invariant under $v \rightarrow v + \const$ (generated by
$\partial/\partial v$) and also under the transformation defined by
equation (\ref{eqn:limit}) (generated by $v \partial/\partial v - r
\partial/\partial r$). These symmetries generate a two-dimensional
non-abelian isometry group $G_2$. The orbits of this group are
two-dimensional if $r \ne 0$ and one-dimensional if $r=0$.

The main point of this paper is to demonstrate that the non-abelian $G_2$ symmetry group is enhanced to a larger $O(2,1)$ symmetry dynamically as a consequence of the Einstein equations.

\subsection{Static black holes}

In the static case, the $O(2,1)$ symmetry can be understood kinematically. For a static black hole, the generator of time translations must be null on the event horizon so the near-horizon geometry is a Killing horizon of this Killing vector field and hence $V \equiv \partial/\partial v$ must be hypersurface-orthogonal: $V \wedge dV=0$. Hence the near-horizon geometry is also static. We then have

\bigskip

\noindent {\bf Lemma 0.} A static near-horizon geometry is locally a warped product of a 2d maximally symmetric space-time with a compact $(D-2)$-manifold.

\bigskip

\noindent {\it Proof.} $V \wedge dV=0$ if, and only if, $dh = 0$ and $dF = Fh$.
Therefore, locally there exists a function $\lambda(x)$ such that $h = d\lambda$ and $F = F_0 \exp(\lambda)$. Now define $R = r \exp(\lambda)$. In coordinates $(v,R,x)$ the near-horizon geometry is
\be
 ds^2 = e^{-\lambda(x)} \left( F_0 R^2 dv^2 + 2 dvdR \right)  + \gamma_{ab}(x) dx^a dx^b.
\ee The terms in brackets describe a 2d maximally symmetric
spacetime:  de Sitter if $F_0>0$, Minkowski if $F_0=0$, and
anti-de Sitter if $F_0 < 0$. $\gamma_{ab}$ is the metric on a spatial cross-section of the horizon, which, for a black hole, is necessarily compact.

\bigskip

The word ``locally" can be deleted if the horizon is simply
connected. For a static black hole $V$ must be timelike outside the
horizon. After taking the near-horizon limit, this gives $F_0 \le 0$ so the de Sitter case is excluded. So
in general, the near-horizon geometry of a static extremal black
hole is locally a warped product of $AdS_2$ or $R^{1,1}$ with a
compact $(D-2)$-manifold. Hence there is a local $O(2,1)$ symmetry
if $F_0 < 0$ and a local 2d Poincar\'e symmetry if $F_0=0$. The
orbits of these isometry groups are 2-dimensional. The symmetry is
global if the horizon is simply connected.

It is possible for a {\it non-static} black hole to have a static near-horizon geometry. Indeed, this is what happens for supersymmetric black rings, which have near-horizon geometry locally isometric to $AdS_3 \times S^2$ \cite{bpsring}. For these solutions, one finds that $F_0=0$ and hence there is a local Poincar\'e symmetry (it is only local because the horizon is not simply connected). The symmetry acts on the flat slices of $AdS_3$ written in ``horospherical" coordinates. In this case, the Poincar\'e symmetry is obviously a subgroup of a much larger local symmetry group, and, as we shall see later, there is also a (global) $O(2,1)$ symmetry present.
In the next section (and in the Appendix), we shall argue that, for black holes with the same amount of rotational symmetry as known solutions, the existence of a near-horizon Poincar\'e symmetry can {\it only} arise in this way, i.e., there will always be a global $O(2,1)$ symmetry in addition to the local Poincar\'e symmetry.

\subsection{Rotational symmetries}

If a stationary, non-extremal, black hole is {\it rotating}, i.e., if the stationary Killing field is not null on the event horizon, then it must be axisymmetric, i.e., it must admit a rotational $U(1)$ symmetry \cite{hawkingellis,Wald}.
Assuming that this is also true in the extremal case, the near-horizon metric (\ref{eqn:nearhor}) must also admit a $U(1)$ symmetry. Hence, on the basis of what has been proved for general black holes, we can expect an extremal rotating black hole to possess a near-horizon $G_2 \times U(1)$ symmetry.

For $D=4$, the $G_2 \times U(1)$ symmetry implies that the near-horizon geometry is cohomogeneity-1. It turns out that the same is true for all {\it known} extremal black holes in $D=5$. The reason is that all such black holes admit {\it two} rotational symmetries. It is not known whether this should be true in general, or whether it is an ``accidental" property of the known solutions. In any case, in $D=5$ we are going to restrict attention to black holes for which this is true. Hence we assume that there is a $U(1)^2$ rotational symmetry, and therefore the near-horizon geometry has $G_2 \times U(1)^2$ isometry group whose generic orbits are 4-dimensional, so the near-horizon geometry is cohomogeneity-1.

For the sake of generality, we shall consider a $D$-dimensional near-horizon geometry with a $G_2 \times U(1)^{D-3}$ isometry whose generic orbits are $(D-1)$-dimensional. For $D>5$, some of the $U(1)$ factors must be translational, rather than rotational, symmetries as the rotation group $SO(D-1)$ admits a $U(1)^{D-3}$ subgroup only for $D=4,5$. The most natural interpretation for $D>5$ is that we are considering a black brane rather than a black hole, with some worldvolume directions wrapped on a torus to give a black hole after reduction to $D=4$ or $D=5$.

If $D=4$ then the only (compact) horizon topologies consistent with the existence of a global rotational Killing field are $S^2$ and $T^2$. If $D=5$ then then only possibilities consistent with two rotational Killing fields are $S^3$ (or a quotient), $S^1 \times S^2$ and $T^3$ \cite{gowdy}.

The existence of the $U(1)^{D-3}$ symmetry allows us to introduce coordinates $x^a = (\rho,x^i)$, $i=1\ldots D-3$, such that $\partial/\partial x^i$ are Killing, and
\be
\gamma_{ab} dx^a dx^b = d\rho^2 + \gamma_{ij}(\rho) dx^i dx^j.
\ee
For toroidal topology, $\gamma_{ij}$ is non-degenerate and periodic in $\rho$. For non-toroidal topology, the range of $\rho$ is a finite interval and $\gamma_{ij}$ degenerates at the endpoints of this interval, where one of the Killing fields vanishes. For $S^2$ or $S^1 \times S^2$ topology, it is the same Killing field that vanishes at each endpoint but for $S^3$ topology, it is a different Killing field at each endpoint \cite{gowdy}.

Define a positive function $\Gamma(\rho)$ by
\be
 h_{\rho} = - \frac{\Gamma'}{\Gamma},
\ee
and functions $k_i(\rho)$ by
\be
 h_i = \Gamma^{-1} k_i.
\ee
We can now perform a coordinate change $r \rightarrow \Gamma(\rho) r$ to bring the near-horizon metric to the form
\be
\label{NH}
 ds^2 = r^2 A(\rho) dv^2 + 2 \Gamma(\rho) dv dr + d\rho^2 + \gamma_{ij}(\rho) \left( dx^i + k^i(\rho) r dv \right)\left( dx^j + k^j(\rho) r dv \right),
\ee
where $k^i \equiv \gamma^{ij} k_j$. We are now ready for

\bigskip

\noindent {\bf Lemma 1}. Consider a near-horizon geometry with symmetry $G_2 \times U(1)^{D-3}$. Introduce coordinates $(v,r,\rho,x^i)$ as above. If the $\rho i$ and $\rho v$ components of the Ricci tensor vanish then
$k^i$ is constant and $A(\rho) = A_0 \Gamma(\rho)$ for some constant
$A_0$. The near-horizon metric is
\be
\label{enhancedNH}
 ds^2 = \Gamma(\rho) \left[ A_0 r^2 dv^2 + 2 dv dr \right] + d\rho^2 + \gamma_{ij}(\rho) \left(  dx^i + k^i r dv \right)\left( dx^j + k^j r dv
 \right).
\ee

\bigskip

\noindent {\it Proof}. Explicit calculation gives (a prime denotes a derivative with respect to $\rho$)
\be
 R_{\rho i } = \frac{1}{2\Gamma} \gamma_{ij} \left( k^j \right)',
\ee
\be
 R_{\rho v} = \frac{r}{\Gamma} \left[A' - \frac{\Gamma'}{\Gamma} A + \left(k^i \right)' k_i \right].
\ee Hence $R_{\rho i}=0$ implies $(k^i)'=0$ and then $R_{\rho v}=0$
implies $A = A_0 \Gamma$.

\bigskip

The part of the metric (\ref{enhancedNH}) in square brackets is the
metric of a 2d maximally symmetric space-time $M_2$: de Sitter if
$A_0>0$, Minkowski if $A_0=0$ and anti-de Sitter if $A_0<0$. The
next lemma shows that all symmetries of this 2d space-time extend to
symmetries of the full near-horizon metric (\ref{enhancedNH}).
\bigskip

\noindent {\bf Lemma 2.} The metric (\ref{enhancedNH}) has an
isometry group $\hat{G}_3 \times U(1)^{D-3}$ where the 3-dimensional
group $\hat{G}_3$ is the 2d Poincar\'e group if $A_0=0$ or $O(2,1)$
if $A_0 \ne 0$. The orbits of $\hat{G}_3$ are 3-dimensional if $k^i
\ne 0$ and 2-dimensional if $k^i=0$.
\bigskip

\noindent {\it Proof.} $M_2$ has a 3-dimensional isometry group $\hat{G}_3$.
We need to show that these isometries extend to the rest of the metric. This is trivial if $k^i=0$ (in which case the orbits of $\hat{G}_3$ are obviously 2-dimensional), so assume $k^i \ne 0$. The volume form of $M_2$ is $dr
\wedge dv = d (rdv)$. This volume form is invariant up to a sign under $\hat{G}_3$. Hence under an isometry in $\hat{G}_3$ we must have $rdv
\rightarrow \pm( rdv + d\phi)$ for some function $\phi(v,r)$. Since $k^i$
is constant, we can compensate for this shift by a $U(1)$
transformation $x^i \rightarrow \pm(x^i - k^i \phi(v,r))$, so the full
metric is invariant under an $\hat{G}_3 \times U(1)^{D-3}$ symmetry. The orbits of $\hat{G}_3$ are 3-dimensional because of this shift in $x^i$.

\bigskip

We shall prove below (and in the Appendix) that the $A_0 \geq 0$
case can be ruled out subject to the additional assumptions on the
theory listed in Theorem 1. Therefore we are mainly interested in
$A_0 < 0$.

It will be useful to have explicit expressions for the discrete
symmetries of $AdS_2$. To do this we transform the $AdS_2$ to global
coordinates, in such a way to make the enhancement of symmetry
manifest in the full near-horizon metric. This can be achieved by
the transformation $(v,r,x^i) \to (T,Y, \chi^i)$ defined by:
 \bea \label{ads2} \qquad\left( \begin{array}{c} r \\ g^2vr \end{array} \right)= \left(\begin{array}{c} -Y + g^{-1}(1+g^2Y^2)^{1/2} \sin(g T)
\\ {(1+g^2Y^2)^{1/2}\cos(gT)-1}\end{array} \right)
\eea and \be \label{eqn:xcoord} dx^i+k^irdv=d\chi^i+k^iYdT, \ee
where for clarity we have written $A_0=-g^2$ since we are concerned with $A_0 \le 0$ (these equations are also valid for Poincar\'e symmetry ($A_0=0$) if one takes the limit $g \rightarrow 0$ with $T,Y$ held fixed).
Note that equation (\ref{eqn:xcoord}) is integrable since $dv \wedge
dr= dT \wedge dY$, and also that $\partial / \partial x^i= \partial / \partial \chi^i$. In these coordinates the near-horizon geometry is
\be
 ds^2 = \Gamma(\rho) \left[ - \left( 1 +g^2 Y^2 \right) dT^2 + \left( 1 +g^2 Y^2 \right)^{-1} dY^2 \right] + d\rho^2 + \gamma_{ij}(\rho) \left(  d\chi^i + k^i Y dT \right)\left( d\chi^j + k^j Y dT \right).
\ee
It is clear that this near-horizon metric exhibits the discrete isometries
\be \label{eqn:discrete}
 P_1: (T,\chi^i) \rightarrow (-T,-\chi^i), \qquad P_2 : (T,Y) \to (-T-Y)
\ee which are inherited from the discrete T and PT isometries of $AdS_2$ respectively. $P_1$ is in $O(2,1)$ but not $SO(2,1)$. $P_2$ is in $SO(2,1)$ but not continuously connected to the identity.

We end this section by examining when the near-horizon geometry (\ref{enhancedNH}) is {\it static}. This occurs if, and only if (i) $k=0$ or (ii) $k^2=-A_0 \Gamma$ with $A_0<0$, and $k_i = \Gamma \bar{k}_i$ where $\bar{k}_i$ is constant. In case (i) it is obvious that Lemma 2 is a special case of Lemma 0.
Case (ii) is more interesting. In this case, we can choose our coordinates $x^i$ so that $k
= g \partial/\partial x^1$ and (by shifting $x^1$ if necessary)
$\bar{k} \propto dx^1$. Split the coordinates as $x^i=(x^1,x^A)$,
$A=2 \ldots D-3$ (for $D \geq 5$). Then we have $\gamma_{11} =k^2/g^2=
\Gamma$ and $\gamma_{1A}=0$. Hence $k=g\Gamma dx^1$. The metric
is \be \label{ads3NH}
 ds^2 = \Gamma(\rho) \left[ -g^2r^2 dv^2 + 2 dv dr +(dx^1+ g r dv)^2 \right] + d\rho^2 + \gamma_{AB}(\rho) dx^A dx^B,
\ee
The metric in square brackets is locally isometric to $AdS_3$. Hence in this case, the near-horizon geometry has local isometry group $O(2,2) \times U(1)^{D-4}$, where $O(2,2)$ has 3d orbits.
$O(2,2)$ is only local because $x^1$ must be identified for the horizon to be compact. Globally, this breaks $O(2,2) \sim O(2,1) \times O(2,1)$ to the $O(2,1) \times U(1)$ guaranteed by Lemma 2.

\subsection{General second order theory}{\label{sec:gentheory}}

As mentioned above, if $D>5$ then some of the Killing directions must parameterize Kaluza-Klein directions so given a theory in $D>5$ dimensions we can work in a dimensionally reduced theory with $D=4$ or $D=5$. We assume that this is of the type described in the Introduction. We can now present the proof of Theorem 1 stated in the Introduction. The method is first to show that a near-horizon solution of this theory will satisfy the assumptions of Lemma 1, and hence (Lemma 2) the metric will possess enhanced isometry group $\hat{G}_3$. Then we show that the other fields also exhibit symmetry enhancement, although only with respect to the subgroup $G_3$ of $\hat{G}_3$ that preserves orientation in $M_2$.

To satisfy the assumptions of Lemma 1, we need to show that $T_{\rho
i } = T_{\rho v} = 0$ in the near-horizon limit for any extremal
black hole solution of this theory, where $T_{\mu\nu}$ is the
energy-momentum tensor. We assume that the matter fields are
invariant with respect to the Killing fields $\partial/\partial v$
and $\partial/\partial x^i$. Hence, after taking the near-horizon
limit, the scalar fields are functions of $\rho$ only. The scalar
kinetic and potential terms make a vanishing contribution to
$T_{\rho i}$ since $g_{AB}(\phi) \partial_\rho \phi^A \partial_i
\phi^B$ and $g_{\rho i}$ are zero. Similarly there is no scalar
field contribution to $T_{\rho v}$.

Turning to the vector fields, we first make use of a standard
result: if $X$ and $Y$ are commuting Killing vector fields that
preserve a closed 2-form $F$ then $F_{\mu\nu}X^\mu Y^\nu$ is constant. Take
$X$ to be a rotational symmetry. For conventional asymptotic
boundary conditions (e.g. asymptotically flat or asymptotically
anti-de Sitter), $X$ must vanish somewhere in the full black hole
space-time (on the ``axis" of rotational symmetry). Hence $F_{\mu
\nu}X ^\mu Y^\nu \equiv 0$.  Taking $X=\partial/\partial x^i$, and $Y=\partial/\partial x^j$ or $\partial/\partial v$, we conclude that $F^I_{ij}$ and $F^I_{vi}$ must vanish. Using this, the near-horizon limit of the Maxwell field
must have the form\footnote{
The near-horizon limit eliminates any $r$-component of $F^I$ except for $F^I_{vr}$.}
\be \label{eqn:nhMaxwell}
 F^I = F^I_{vr}(\rho) dv \wedge dr + r \tilde{F}^I_{v \rho}(\rho) dv \wedge d\rho + F^I_{\rho i}(\rho) d\rho \wedge dx^i.
\ee
Imposing the Bianchi identity $dF^I=0$ implies $(F^I_{vr})'=\tilde{F}^I_{v\rho}$. One then finds (using the metric (\ref{NH})) that
\be
 \left[d \left( g_{IJ}(\phi) \star F^J \right) \right]_{rvi_1 \ldots i_{D-3}} = \sqrt{\gamma} g_{IJ} (\phi)   \left( \tilde{F}^I_{v \rho} + k^i (\rho) F^I_{\rho i} \right).
\ee
The equation of motion for $A^I$ says that this must be proportional to $d(h_{IJ}(\phi) F^J)_{rvi_1}$ if $D=4$ or to $C_{IJK}(F^J \wedge F^K)_{rvi_1 i_2}$ if $D=5$ but it is easy to see that both of these terms vanish. Hence the Maxwell equation implies that
\be
\label{Delta}
 \tilde{F}^I_{v \rho} =- k^i(\rho) F^I_{\rho i}.
\ee
Substituting this back into (\ref{eqn:nhMaxwell}) gives
\be
\label{eqn:nhMaxwell2}
 F^I = F^I_{vr}(\rho) dv \wedge dr + F^I_{\rho i} (\rho) d\rho \wedge \left( dx^i +   k^i(\rho) r dv \right).
\ee It is then easy to see that the Maxwell fields makes a vanishing
contribution to $T_{\rho i}$ and $T_{\rho v}$. Hence we have
satisfied the conditions of Lemma 1 so we must have $A=A_0 \Gamma$ and $k^i$ is constant. Therefore, from Lemma 2, the metric exhibits an enhanced isometry group $\hat{G}_3$.

Converting (\ref{eqn:nhMaxwell2}) to global coordinates gives
\be
 F^I = F^I_{TY}(\rho)dT \wedge dY + F^I_{\rho i}(\rho)d\rho \wedge \left( d\chi^i + k^i Y dT \right).
\ee
It is now obvious that the Maxwell fields inherit all the {\it continuous} enhanced symmetries of the metric, as well as the discrete symmetry $P_2$ of (\ref{eqn:discrete}). However, under the discrete symmetry $P_1$, we have
\be
\label{eqn:discretemaxwell}
 F^I \rightarrow - F^I.
\ee
Hence, although the metric and scalars are invariant under $\hat{G}_3$, the Maxwell fields are only invariant under the subgroup of $\hat{G}_3$ that preserves orientation in $M_2$, which we shall denote as $G_3$. If $A_0 <0$ then $G_3 = SO(2,1)$.

We can now rule out the $A_0 \geq 0$ case, which we analyze in the
Appendix (note this includes the Poincar\'e symmetric case $A_0=0$).
If the matrices $f_{AB}$ and $g_{IJ}$ occurring in the scalar and
vector kinetic terms are positive-definite (as they will be for
sensible theories) and the scalar potential is non-positive, then
the argument in the Appendix proves that $A_0 \leq 0$. This rules
out $A_0>0$. Further, if $A_0=0$ then $k^i=0$, the Maxwell fields
must vanish, the scalars must take constant values such that the
potential vanishes (if the potential is strictly negative then this
is already a contradiction), and the near-horizon geometry must be
flat: a direct product of $R^{1,1}$ and $T^{D-2}$. Hence this case
can only arise for toroidal horizon topology, and is therefore
excluded if we assume that the horizon is non-toroidal. This
concludes the proof of Theorem 1.

Finally, we return to the special case in which the near-horizon metric has $AdS_3$ symmetry (equation (\ref{ads3NH})). An obvious question is whether the Maxwell fields in the general theory considered here also inherit the
symmetries of $AdS_3$. We will focus on the $D=5$ case. The
$D=5$ Einstein equation is:
\begin{equation}
R_{\mu\nu} = \frac{1}{2}f_{AB}\partial_{\mu}\phi^{A}\partial_{\nu}\phi^{B} + \frac{1}{3}V(\phi)g_{\mu\nu} + \frac{1}{2}g_{IJ}F^{I}_{\mu \alpha}F_{\nu}^{J\alpha} - \frac{1}{12}g_{IJ}F_{\alpha \beta}^{I}F^{J\alpha\beta}g_{\mu\nu}.
\end{equation}
For a metric of the form (\ref{ads3NH}), $R_{vv}=0$ automatically, so the $(vv)$ component of the Einstein equation becomes
\begin{equation}
g^2 r^2\left(F_{\rho 1}^{I}F_{\rho 1}^{J}g_{IJ} + F_{vr}^{I}F_{vr}^{J}g_{IJ}\Gamma^{-1}\right) = 0
\end{equation}
and hence, assuming $g_{IJ}$ is positive definite, $F^I_{vr}=0$ and $F^I_{\rho 1}=0$. Therefore the Maxwell
field simplifies to \be F^I= F_{\rho 2}^I d\rho \wedge dx^2 \ee
which is manifestly $O(2,2)\times U(1)$ invariant. The scalars are
trivially invariant under this symmetry, and thus we learn that in
this special case the full solution must be $O(2,2) \times U(1)$
invariant.

In this special case with $AdS_3$ symmetry, the horizon topology must be $S^1 \times S^2$. The near-horizon geometry is generically a warped product of $AdS_3$ and $S^2$ (with the warp factor a function of the polar angle on $S^2$). However, if $\Gamma$ is a constant, then one can argue (using the equations of motion) that the near-horizon geometry is a {\it direct} product of (locally) $AdS_3$
and $S^2$ with constant scalars. The near-horizon of the supersymmetric black ring \cite{bpsring} is in
this class. In a recent investigation of the existence of asymptotically $AdS_5$ supersymmetric black rings, we found a near-horizon geometry with $AdS_3$ symmetry and non-trivial warping but it was not possible to eliminate a conical singularity from the $S^2$ \cite{KLR}.

\subsection{Higher derivative corrections}

\label{sec:hd}

Much of the recent interest in the attractor mechanism derives from its validity in the presence of higher-derivative terms. It is therefore of interest to examine whether such terms affect our result. In this section we will prove Theorem 2 stated in the Introduction. The following lemma will prove useful:

\bigskip

\noindent {\bf Lemma 3.} Consider the $O(2,1) \times U(1)^{D-3}$-symmetric near-horizon space-time (\ref{enhancedNH}). Let $J$ be a conserved current invariant under $SO(2,1) \times U(1)^{D-3}$. Assume that $m_i \equiv \partial/\partial x^i$ vanishes somewhere in the near-horizon geometry for some $i$ (as will be the case if the horizon topology is non-toroidal). Then $J^\rho=0$.

\bigskip

\noindent {\it Proof.} $SO(2,1) \times U(1)^{D-3}$ symmetry implies that
\be
 J = J^\rho(\rho) \frac{\partial}{\partial \rho} + J^i (\rho) \frac{\partial}{\partial x^i}.
\ee
Plugging this into the conservation equation in the background (\ref{enhancedNH}) gives
\be
 0 = \partial_\mu \left( \Gamma \sqrt{\gamma} J^\mu \right) = \frac{d}{d\rho} \left(  \Gamma \sqrt{\gamma} J^\rho \right),
\ee
where $\gamma = \det \gamma_{ij}$. Hence
\be
 J^\rho = \frac{j}{\Gamma \sqrt{\gamma}},
\ee
where $j$ is a constant. But now consider
\be
 \star \left( m_1 \wedge m_2 \wedge \ldots \wedge m_{D-3} \right) = \Gamma \sqrt{\gamma} dv \wedge dr \wedge d\rho.
\ee
Let $i_J$ denote the operation of contracting $J$ with the first index of a $p$-form. Then
\be
 i_J \star \left( m_1 \wedge m_2 \wedge \ldots \wedge m_{D-3} \right) = j dv \wedge dr = j dT \wedge dY.
\ee
Now evaluate the LHS where $m_i$ vanishes to conclude that $j=0$ and the result follows.

\bigskip

Now we can examine higher-derivative corrections. Assume that we are dealing with a theory with an action of the form (\ref{eqn:hdaction}) described in Theorem 2.
Varying this action will lead to an Einstein equation of the form
\be
\label{eqn:hdeinstein}
 R_{\mu \nu}-\frac{1}{2}R g_{\mu\nu} = T_{\mu\nu}+\sum_{m \ge 1}\lambda^m H^{(m)}_{\mu \nu},
\ee
where $T_{\mu\nu}$ is the energy momentum tensor of the 2-derivative part of the action, and $H^{(m)}$ is conserved. The Maxwell equations will take the form
\be
 \nabla_\mu \left( g_{IJ} (\phi) F^{J \mu \nu} \right) + S^\nu = \sum_{m \ge 1} \lambda^m K^{(m)\nu},
\ee
where $S^\nu$ is the contribution to the equation of motion from the term $S_{\rm top}$ in the 2-derivative part of the action and $K^{(m)}$ is conserved.

Consider a general $G_2 \times U(1)^{D-3}$ invariant near-horizon solution of these equations of motion. Assume that it is analytic in $\lambda$, and possesses the enhanced $SO(2,1) \times U(1)^{D-3}$ symmetry for $\lambda=0$ (as follows from our analysis of the 2-derivative theory above). We shall present an inductive argument that the solution must be $SO(2,1) \times U(1)^{D-3}$ invariant to all orders in $\lambda$.

Introduce the coordinates $(v,r,\rho,x^i)$ as described above. Assume, inductively, that the solution admits enhanced symmetry up to order $\lambda^n$, in other words we have $g_{\mu \nu} = \bar{g}_{\mu \nu} + \lambda^{n+1} h_{\mu \nu}$ where the components of $\bar{g}_{\mu\nu}$ are polynomials of degree $n$ in $\lambda$, and $\bar{g}_{\mu\nu}$ exhibits symmetry enhancement. Do the same for the other fields.
We then have
\be
 H^{(m)} [g,\phi^A,F^I] = H^{(m)}[\bar{g},\bar{\phi^A},\bar{F}^I] + {\cal O}(\lambda^{n+1}).
\ee
Hence
\be
 R_{\mu\nu}[g]-\frac{1}{2}g_{\mu\nu}R[g] = T_{\mu\nu}[g,\phi^A,F^I]+ \sum_{m \ge 1}\lambda^m H^{(m)}_{\mu \nu}[\bar{g},\bar{\phi^A},\bar{F}^I] + {\cal O}(\lambda^{n+2}).
\ee
$H^{(m)}[\bar{g},\bar{\phi}^A,\bar{F}^I]$ must be invariant under $SO(2,1) \times U(1)^{D-3}$ since it is built from the $SO(2,1) \times U(1)^{D-3}$-invariant objects $\bar{g}$, $\bar{\phi}^A$ and $\bar{F}^I$. Furthermore, we know that $H^{(m)}[\bar{g},\ldots]$ must be conserved with respect to the metric $\bar{g}$. Hence $J_\mu \equiv H^{(m)}_{\mu\nu}[\bar{g},\ldots] m_i^\nu$ is a $SO(2,1) \times U(1)^{D-3}$-invariant conserved current with respect to $\bar{g}$ so, from Lemma 3 we must have $J_\rho = 0$. It follows that $H^{(m)}_{\rho i}[\bar{g},\ldots]=0$. $SO(2,1) \times U(1)^{D-3}$ symmetry implies\footnote{
A symmetric tensor invariant under $SO(2,1) \times U(1)^{D-3}$ must have the form $S_{\mu\nu} dx^\mu dx^\nu = S_0(\rho) \left( A_0 r^2 dv^2 + 2 dv dr \right) + S_1(\rho) d\rho^2 + 2 S_i(\rho) d\rho \left( dx^i + k^i r dv \right) + S_{ij}(\rho) \left( dx^i + k^i r dv \right)\left( dx^j + k^j r dv \right)$. }
$H^{(m)}_{\rho v}=k^i r H^{(m)}_{\rho i}$, hence $H^{(m)}_{\rho v}[\bar{g},\ldots]=0$ too. Therefore the higher derivative correction to the $\rho i$ and $\rho v$ components of the Einstein equation is of order $n+2$ in $\lambda$.

The scalar fields exhibit enhanced symmetry trivially since they are functions only of $\rho$ in the near-horizon limit. Turning to the vectors, we have
\be
 K^{(m)}[g,F^I,\phi^A] = K^{(m)}[\bar{g},\bar{F}^I,\bar{\phi}^A] + {\cal  O}(\lambda^{n+1}).
\ee
Our induction hypothesis implies that the vector $K^{(m)}[\bar{g},\bar{F}^I,\bar{\phi}^A]$ is invariant under $SO(2,1) \times U(1)^{D-3}$. It is also conserved with respect to $\bar{g}$. Hence Lemma 3 implies that $K^{(m)}_\rho[\bar{g},\bar{F}^I,\bar{\phi}^A]=0$. Therefore the higher-derivative correction to the $\rho$-component of the Maxwell equation is of order $n+2$ in $\lambda$.

We can now repeat the argument of section (\ref{sec:gentheory}). The only difference is the ${\cal O}(\lambda^{n+2})$ corrections to the Einstein and Maxwell equations. The result is that $R_{\rho i}$ and $R_{\rho v}$ are ${\cal O}(\lambda^{n+2})$. We conclude that $k^i$ is constant, and $A=A_0 \Gamma$, to order $n+1$ in $\lambda$, and therefore $g$ and $F^I$ exhibit symmetry enhancement to order $n+1$ in $\lambda$.

Having assumed that the fields exhibit enhanced symmetry to order $n$ we see that they must have enhanced symmetry to order $n+1$. We have already proved the result for $n=0$ and hence the result must be valid for all $n$ by induction.

\section{Examples}

\subsection{Determining near-horizon geometries}

In this section we will analyse various examples of near-horizon
limits of known five-dimensional extremal black holes. These will illustrate
some of our general results. We will focus on near-horizon limits of
cohomogeneity-2 black holes as these are the most complicated known
examples.


The first step in determining a near-horizon geometry is to
introduce coordinates regular at the horizon. Rather than giving
details for each solution, we shall explain here how to do this for
a general class of metrics which encompasses all of the solutions we
are interested in. We assume that the black hole metric takes the
cohomogeneity-2 form \be \label{standard}
 ds^2 = g_{tt}(R,x)dt^2+
2g_{ti}(R,x)dtd\Phi^i + g_{RR}(R,x)dR^2+ g_{xx}(R,x)dx^2+
g_{ij}(R,x)d\Phi^id\Phi^j \ee where $R$ is a ``radial" coordinate such
that $R=0$ is the event horizon, and $x$ is a ``polar angle" on the surfaces of constant $t$ and $R$.
All known rotating black hole solutions (including black rings) take the above form.
By shifting $\Phi^i$ by appropriate (constant) multiples of $t$, we can ensure that the coordinates are co-rotating, i.e., $\xi \equiv \partial_t$ is null on the horizon. For known extremal black hole solutions we have
\be
\label{assume}
 g_{ti}= f_i(x)R +\mathcal{O}(R^2), \qquad
 g_{tt}=f_t(x)R^2+\mathcal{O}(R^3), \qquad g_{RR}=f_R(x)R^{-2}
 +\mathcal{O}(R^{-1})
\ee
for certain functions $f_{\mu}(x)$.

The above coordinates are not regular on the horizon.
Therefore we define new coordinates $(v,r,\phi^i)$ by
\be
R=r, \qquad dt= dv+a(r)dr, \qquad d\Phi^i=
d\phi^i+b^i(r)dr \ee where: \be a(r)=
\frac{a_0}{r^2}+\frac{a_1}{r}, \qquad b^i(r)=
\frac{b_0^i}{r}.\ee
The constants are chosen to make the metric and its inverse analytic at $r=0$.
The near-horizon limit is then defined by $v \to v/\epsilon$, $r \to
\epsilon r$ and $\epsilon \to 0$, and let us denote the limiting metric by $\hat{g}_{\mu\nu}$.
The following components are easily obtained since they are not affected
by the transformation to the new coordinates: \be
\hat{g}_{vi}=f_i(x)r, \qquad \hat{g}_{vv}=f_t(x)r^2, \qquad \hat{g}_{ij}=g_{ij}(0,x),
\qquad \hat{g}_{xx}= g_{xx}(0,x). \ee
Comparing to our standard form for
the near-horizon (\ref{NH}) allows one to identify $x^i=\phi^i$, $d\rho^2 = \hat{g}_{xx}(x) dx^2$, and
\be \label{data}
k_i=f_i(x), \qquad A+ k^ik_i=f_t(x), \qquad \gamma_{ij}=\hat{g}_{ij}.
\ee
The $vr$
component of the metric is $g_{vr}=a(r)g_{tt}+b^i(r)g_{ti}$. Taking the near-horizon limit gives: \be \label{Gamma}
\Gamma= \hat{g}_{vr}=a_0f_t(x)+b^i_0f_i(x). \ee
Hence the near-horizon solution is fully-determined once we know the
constants $a_0,b^i_0$. These can be obtained from regularity of $g_{ri}$ and $g_{rr}$. Absence of a $1/r$ term in $g_{ri}$ implies that
\be \label{bconstant}
b^i_0=-a_0\gamma^{ij}f_j(x)= -a_0k^i. \ee
This implies that $k^i$ must be constant. However, this equation is only consistent if $\gamma^{ij} f_j$ is constant. Fortunately this turns out to be true for known solutions. Of course, this is not an accident: we are discussing solutions of the Einstein equation, and we have seen in previous sections that the Einstein equation implies that $k^i$ must be constant.

Absence of a $1/r^2$ term in $g_{rr}$ implies that\footnote{
There will also be a $1/r$ divergence in $g_{rr}$. Eliminating this determines $a_1$. However, we do not need to know $a_1$ to determine the near-horizon geometry since a $dr^2/r$ term vanishes in the near-horizon limit. }
\be a_0^2=
\frac{f_R(x)}{k^ik_i-f_t(x)}. \ee
This determines $a_0$ up to a
sign, and hence $b^i_0$ and $\Gamma$ are also determined up to
the same sign. We choose the sign such that $\Gamma>0$, which ensures that we are dealing with a future, rather than past, horizon. The final
piece of near-horizon data, $A$ is then determined from (\ref{data}).
Combining this with (\ref{Gamma}) and (\ref{bconstant}) gives:
\be \label{A} A=a_0^{-1}\Gamma.
\ee
Equations (\ref{bconstant}) and (\ref{A}) imply that the near-horizon
metric is of the form (\ref{enhancedNH}) with $A_0=a_0^{-1}$. Thus
we see the enhancement of symmetry, which we derived earlier using
more general arguments.


\subsection{$S^3$ topology black holes}

The simplest cohomogeneity-2 black hole with an $S^3$ topology
horizon is the doubly spinning Myers-Perry solution \cite{MyersPerry} (with $a \neq
b$, where $a,b$ are the rotation parameters). Using the general
formalism developed in the previous section we can calculate the
near-horizon limit of the extremal doubly spinning Myers-Perry black
hole with $a \neq b$. Without loss of generality we choose $0<a<b$.
After a tedious calculation, we find that the near horizon limit can
be written in the form (\ref{enhancedNH}) where:
\bea
\label{MPNH}\gamma_{ab}dx^a dx^b &=&\sigma(\theta)^2d\theta^2
\nonumber \\ &+&
\frac{(a+b)^2}{\sigma(\theta)^2}\left(b\cos^2\theta(b+a\cos^2\theta)d{\psi}^2
+ 2r_{0}^2\cos^2\theta\sin^2\theta d{\phi}d{\psi} +
a\sin^2\theta(a+b\sin^2\theta)d{\phi}^2 \right) \nonumber \\
 \Gamma &=& \frac{\sigma(\theta)^2}{(a+b)^2}, \qquad A_0=-\frac{4}{(a+b)^2}, \qquad k^i
\frac{\partial}{\partial x^{i}} =
\frac{2r_{0}}{b(a+b)^2}\frac{\partial}{\partial {\psi}} +
\frac{2r_{0}}{a(a+b)^2}\frac{\partial}{\partial {\phi}}\eea where
the coordinates on the horizon are $x^a=(\theta, \phi, \psi)$ and
$\sigma(\theta)^2= r_0^2+a^2\cos^2\theta+b^2\sin^2\theta$, $r_0^2=ab$.

\subsection{ $S^1 \times S^2$ topology black holes}

{\it Vacuum solutions.} There are two known vacuum solutions with horizon topology $S^1
\times S^2$. These are the (boosted) Kerr string and the black ring \cite{ring,Pomeransky}\footnote{Only the black ring is asymptotically flat.}. The 3-parameter black ring solution admits an extremal limit \cite{Pomeransky}. In this section, we will use the formalism described above to show that the near-horizon geometry of an extremal black ring is globally isometric to that of an extremal Kerr string.

Let us begin with the boosted Kerr string. To construct this
solution one takes the direct product Kerr $\times S^1$, and write
the metric on $S^1$ as $dz^2$. Now perform a boost $(t,z) \to (\cosh
\beta t+ \sinh \beta z, \sinh \beta t+ \cosh \beta z)$, where $t$ is
the time coordinate in which the Kerr metric is at rest at infinity.
After taking the extremal limit, we find that the near-horizon geometry has the expected $O(2,1) \times U(1)^2$ symmetry. The near-horizon data is: \bea
\gamma_{ab} dx^a dx^b &=& a^2(1+\cos^2\theta)d\theta^2+ \frac{4a^2
\sin^2\theta}{1+\cos^2\theta} \left( d\Phi + \frac{\sinh\beta}{2a}
dz \right)^2 + \cosh^2\beta dz^2, \\ \Gamma &=&
\frac{1+\cos^2\theta}{2 \cosh\beta}, \qquad A_0=-\frac{1}{2a^2
\cosh\beta}, \qquad k^i\partial_i = \frac{1}{2a^2 \cosh\beta}
\partial_{\Phi} \eea
where $(\theta, z,\Phi)$ are the coordinates on the horizon\footnote{This $z$ is not the same as the initial one: it has suffered from two shifts, a constant
one in the $t$ direction (to go to co-rotating coordinates) and a singular one in the $r$ direction (to go to coordinates regular at the horizon).} and $a$ is the Kerr angular momentum parameter. Note that
$\Phi$ has period $2\pi$, whereas $z$ may have any period $\Delta z$. This near-horizon solution thus depends on three parameters $(a,\beta,\Delta
z)$.

Let us now turn to the extremal limit of the recently discovered vacuum
black ring with two independent angular momenta~\cite{Pomeransky}.
The solution has three parameters $(k,\lambda,\nu)$ and the extremal
limit is reached when $\nu = \lambda^2/4$, and $0 \leq \lambda \leq
2$. After an involved calculation, we find the near horizon can be
written in the form~(\ref{enhancedNH}) with\footnote{ Note that
unlike in \cite{Pomeransky} we work with a mostly positive
signature, and we call $\psi$ the angle along the $S^1$ of the
ring.}
\begin{eqnarray} \label{dringNH} \nonumber
\gamma_{ab}dx^a dx^b &=& \frac{8\lambda^2k^2 H(x)}{(\lambda x+2)^4(1-x^2)(4-\lambda^2)} dx^2 \\ &+& \frac{32\lambda^2k^2(1-x^2)}{(4-\lambda^2)H(x)} \left(d\phi+ \frac{\lambda^2+8\lambda +4}{4\lambda} d\psi\right)^2 + \frac{4(2+\lambda)^2k^2}{(2-\lambda)^2}d{\psi}^2 \\
\Gamma &=&\frac{k\lambda^2H(x)}{2(2+\lambda x)^2(2+\lambda)^2} ,
\qquad A_0= -\frac{(2-\lambda)^2}{16k}, \qquad k^i \partial_i =
\frac{(2-\lambda)^2}{16k} \partial_{\phi}, \nonumber
\end{eqnarray}
where $H(x) = (\lambda^2+4)(1+x^2) + 8\lambda x$, and
$(x,\psi,\phi)$ are coordinates on the horizon such that $ -1 \leq x
\leq 1$ and $\psi,\phi$ both have period $2\pi$.

We can now prove that this 2-parameter near-horizon geometry is globally isometric to a special case of the 3-parameter Kerr string near-horizon geometry. First, it is necessary to rescale the $v$ coordinate of the boosted Kerr string: $v \to Cv$. The following coordinate transformation then proves they are globally equivalent:
\bea \cos\theta = \frac{2x+\lambda}{2+\lambda x}, \qquad
\Phi=\phi+\psi, \qquad z= \frac{\sqrt{2}k(2+\lambda)}{(2-\lambda)}
\psi \eea
provided that the Kerr string parameters are chosen to be:
\be a^2 =
\frac{8k^2\lambda^2}{(4-\lambda^2)^2}, \qquad \sinh^2\beta =1,
\qquad \Delta z =\frac{2\sqrt{2}\pi k (2+\lambda)}{2-\lambda} \qquad
C= \frac{\sqrt{2}k\lambda^2}{(2+\lambda)^2}.
\ee
{\it Dipole rings.} The dipole ring solution of $D=5$ Einstein-Maxwell theory \cite{Emparan} admits an extremal limit even though it rotates in only one plane. We will work in the conventions of~\cite{ringreview}.
After taking the near horizon limit as described above we find the that the resulting solution has $SO(2,1) \times U(1)^2$ symmetry with
\begin{eqnarray} \label{dipoleringNH} \nonumber
\gamma_{ab}dx^a dx^b &=& R^2F(x)H(x)\mu^2 \left[ \frac{dx^2}{1-x^2} +
\frac{1-x^2}{F(x)H(x)^3} d\phi^2 \right] + \frac{R^2\lambda(1+\lambda) H(x)}{\mu (1-\lambda)F(x)}d\psi^{2}   \\
\Gamma &=&  \left(\frac{\mu(1-\lambda)}{\lambda(1+\lambda)}\right)^{1/2}R F(x)H(x),
\qquad A_0= -\frac{(1-\lambda)^{1/2}}{R\left(\lambda(1+\lambda) \mu^{3}\right)^{1/2}}, \\ k^i \partial_i &=&
-\frac{(1-\lambda)^{3/2}}{R\lambda(1+\lambda)^{1/2} } \partial_{\psi}, \nonumber
\end{eqnarray} where $F(x)=1+\lambda x$ and $H(x)=1-\mu x$, with $0 \leq \lambda, \mu <1$. The gauge potential in this case is simply given by:
\be A= \left( \frac{1-\mu}{1+\mu} \right)^{1/2}\frac{\mu R
(1+x)}{H(x)} d\phi \ee

\subsection{Analytic continuation}

We have seen that near-horizon solutions necessarily possess an $SO(2,1)$ symmetry. For non-static black holes, $SO(2,1)$ has 3-dimensional orbits, which have the form of a line, or circle, bundle over $AdS_2$. In this section, we note that near-horizon solutions can sometimes be analytically continued so that $AdS_2$ becomes $S^2$ and $SO(2,1)$ becomes $SU(2)$ acting on a circle bundle over $S^2$ (this bundle is just $S^3$ in the case we shall discuss). This generalizes the analytic continuation relating solutions with $AdS_3$ and $S^3$ symmetries that has been studied in \cite{LPV,Gauntlett}.

Consider first the near-horizon geometry of the extremal dipole ring discussed above.
One can analytically continue this near-horizon geometry to obtain an $SU(2)$-symmetric non-extremal Kaluza-Klein black hole solution. To see this, transform from $(v,r,\psi)$ to the global coordinates $(T,Y,\chi)$ defined earlier (\ref{ads2},
\ref{eqn:xcoord}) and continue $Y \to ig^{-1}\cos\theta$, $\chi \to ik^{\psi}g^{-2}\chi$ and rescale $\phi \to t/R$, $T \to g^{-1}T$, where $(g,k^{\psi})$ are given in~(\ref{dipoleringNH}).  The result is:
\bea
\nonumber ds^2 &=& -\frac{\mu^2(x^2-1)}{(\mu x-1)^2} dt^2 +
\frac{R^2F(x)(\mu
x-1)dx^2}{x^2-1} + R^2\mu^2(\mu x-1) F(x)( d\theta^2 + \sin^2\theta dT^2) \\
&+& R^2\mu^2(1-\lambda^2) \frac{(\mu x-1)}{F(x)} (d\chi + \cos\theta dT)^2, \eea It is easy to see one may write the $(\theta,T,\chi)$ part of the metric in terms of left invariant one-forms on $SU(2)$, so that $\sigma_3 = d\chi + \cos\theta dT$ and $\sigma_1^2+\sigma_2^2 = d\theta^2+\sin^2\theta dT^2$.  Now make
the following coordinate transformation: $ r = R\sqrt{\lambda
\mu}(\mu x-1)$. This gives: \bea ds^2 = -Vdt^2+ \frac{U}{V}dr^2 + U
r^2 (\sigma_1^2+\sigma_2^2) + W \sigma_3^2, \qquad F= d
\left[\frac{\sqrt{r_+r_-}}{r}dt \right] \eea where \bea V=
\frac{(r-r_+)(r-r_-)}{r^2}, \qquad U =\frac{r+r_0}{r}, \qquad W=
\frac{(r_++r_0)(r_-+r_0)}{U} \eea and \be r_{\pm}= -R \sqrt{\lambda
\mu}(1\pm \mu), \qquad r_0= R\sqrt{\lambda \mu}\left( 1+
\frac{\mu}{\lambda} \right). \ee This is the Kaluza-Klein black hole
solution discussed in \cite{IM}.\footnote{In order for this
specetime to be regular on and outside the horizon $r=r_+$ we
require $r_+>0$, $r_+>r_-$ and $r_+>-r_0$. These conditions can be
fulfilled in two ways: (i) $R<0$, $\mu>0$, $\lambda>1$ or (ii)
$R>0$, $\mu<-1$, $\lambda<0$.}

Next we shall show that the near-horizon geometry (\ref{MPNH}) of the extremal Myers-Perry black hole is related by analytic continuation to the non-extremal self-dual Myers-Perry solution, which has metric
\bea
ds^2 &=& -f^2dt^2+g^2 dr^2+ \frac{h^2}{4}(\sigma_3-\Omega dt)^2+ \frac{r^2}{4}(\sigma_1^2+\sigma_2^2), \\
g^{-2} &=& 1-\frac{2M}{r^2}+\frac{2M\alpha^2}{r^4}, \qquad
h^2=r^2\left( 1+\frac{2M\alpha^2}{r^4} \right), \qquad f=
\frac{r}{hg}, \qquad \Omega= \frac{4M\alpha}{r^2h^2} \eea where
$\sigma_i$ are again left-invariant one-forms on $SU(2)$. To do this, we will work backwards from the self-dual Myers-Perry solution.
First we make
the following coordinate change: $r^2=4\Gamma/C^2$, $y=\cos\theta$
where we denote the Euler angles by $(\theta, \tau, \psi)$ so
$\sigma_3= d\psi+ \cos\theta d\tau$ and $\sigma_1^2+\sigma_2^2=
d\theta^2+\sin^2\theta d\tau^2$. Now perform the following
analytic continuation: $y\to iy$, $C^2\to -C^2$ and $\alpha \to
i\alpha$. The resulting metric is: \bea ds^2 = \frac{\Gamma}{C^2}
\left[ -(1+y^2)d\tau^2 + \frac{dy^2}{1+y^2} +
\frac{A(\Gamma)}{\Gamma^2} ( d\psi +yd\tau +\omega(\Gamma) dt)^2
\right] + \frac{\Gamma d\Gamma^2}{4P(\Gamma)} + \frac{4P(\Gamma)
dt^2}{C^2A(\Gamma)} \eea where \bea
P(\Gamma) &=& - \frac{C^2}{4}\Gamma^2- \frac{MC^4}{8} \Gamma + \frac{M\alpha^2C^6}{32}, \\
A(\Gamma) &=& \Gamma^2-\frac{M\alpha^2C^4}{8}, \qquad
\omega(\Gamma)= -\frac{M\alpha C^4}{4A(\Gamma)}. \eea Using the
inverse of the $AdS_2$ coordinate transformation used earlier
(\ref{ads2}, \ref{eqn:xcoord}), with $C=g$, $y=gY$, $\tau=gT$,
$\psi=g^2 \chi$, and letting $x^2=t$, we get the following geometry:
\be \label{MPAC} ds^2 = \Gamma \left[ -C^2r^2dv^2 + 2dvdr +
C^2\frac{A(\Gamma)}{\Gamma^2} ( dx^1 +rdv+
C^{-2}\omega(\Gamma)dx^2)^2  \right] + \frac{ \Gamma
d\Gamma^2}{4P(\Gamma)} + \frac{4P(\Gamma) (dx^2)^2}{C^2A(\Gamma)}.
\ee It is now apparent that this metric looks like a near-horizon
geometry. The corresponding metric on the horizon at $r=0$ is: \be
\label{3sphere} \gamma_{ab}dx^adx^b =
\frac{C^2A(\Gamma)}{\Gamma}(dx^1+C^{-2}\omega(\Gamma)dx^2)^2+ \frac{
\Gamma d\Gamma^2}{4P(\Gamma)} + \frac{4P(\Gamma)
(dx^2)^2}{C^2A(\Gamma)}. \ee In order to prove that this
near-horizon geometry is the near-horizon limit (\ref{MPNH}), it is
necessary that the horizon metric (\ref{3sphere}) can be made
globally regular with $S^3$ topology. Enforcing compactness and
regularity allows us to deduce the required coordinate change to
prove the equivalence of the two metrics. The calculation is similar
to that done in~\cite{KLR} and we omit the details. The explicit
transformations $(\Gamma, x^{i}) \rightarrow (\theta, \psi, \phi)$
are given by:
\begin{equation}
\Gamma = \frac{\sigma(\theta)^2}{(a+b)^2}, \qquad  x^1 = \frac{r_{0}(a+b)^2}{2(a-b)}\left( \psi - \phi \right), \qquad x^2 = \frac{(a+b)}{(a-b)}\left(a\phi - b\psi \right),
\end{equation}  and the parameters are related by
\begin{equation}
C^2  =  \frac{4}{(a+b)^2} \qquad  M = - \frac{(a+b)^2}{2} \qquad
\alpha^2 = ab.
\end{equation} With these identifications it is straightforward to confirm that the two near-horizon
metrics (\ref{MPNH}) and (\ref{MPAC}) are identical (although the mass $M$ has to be taken negative). Notice that the
$(\Gamma,x^i)$ coordinate system is simpler than the ``natural"
coordinates one obtains from taking the near-horizon limit. This is
actually the same coordinate system encountered in~\cite{KLR} for
the near-horizon geometry of a supersymmetric $AdS_5$ doubly
spinning black hole~\cite{Chong}.

The above analytic continuation can be generalised to
other near horizon geometries with spherical topology horizons, such
as Myers-Perry-AdS~\cite{HHT} and charged versions of
this~\cite{Chong,KLR2}. In particular, we find that the near horizon
limit of the Chong {\it at al} supersymmetric $AdS_5$ black
hole~\cite{KLR} analytically continues to the Klemm-Sabra ``time-machine"~\cite{SabraKlemm}. This implies that the number of supersymmetries of the Chong {\it et al} solution is enhanced from two to four in the near horizon limit.

We should point out that analytic continuation does not always lead to a stationary black hole solution. For example, if one starts from the near-horizon Kerr solution then one ends up with a special case (vanishing mass) of the Lorentzian Taub-NUT solution. This has $SO(3) \times U(1)$ symmetry, where $SO(3)$ has 3-dimensional orbits and acts non-trivially on the time coordinate, giving rise to a non-vanishing NUT charge. There is no way of avoiding this in $D=4$ (except for giving up stationarity). In $D=5$ the near-horizon symmetry group is $SO(2,1) \times U(1)^2$. This group has 4-dimensional orbits, but the orbits of $SO(2,1)$ are only 3-dimensional. This gives the possibility of analytically continuing in such a way that the new ``time" direction lives within the surfaces of homogeneity, but is not acted on by $SU(2)$ (or $SO(3)$), thereby avoiding NUT charge. This is precisely what we have done above.

\subsection{The general solution for a vacuum near-horizon geometry}

Lemma 1 tells us that a Ricci flat near-horizon geometry must be of the form (\ref{enhancedNH}). In this section we shall show that one can determine, at least implicitly, the general Ricci-flat solution of this form.
We will assume $A_0 \neq 0$ since otherwise $k=0$ and the geometry is static, which leads to a trivial near-horizon geometry~\cite{crt} (assuming a compact horizon). Our method is motivated by the analytic continuation described above:
continuation of a $SO(2,1) \times U(1)^{D-3}$-invariant vacuum near-horizon solution gives a vacuum solution with symmetry group $\sim SO(3) \times U(1)^{D-3}$. This is the symmetry of a stationary, spherically symmetric solution of $D$-dimensional Kaluza-Klein theory. All such solutions were obtained in \cite{Maison2} using a method introduced in \cite{Maison1}. Hence, by repeating the analysis of \cite{Maison1,Maison2} we can determine the general vacuum near-horizon solution.

We start with a reduction to three dimensions on the $D-3$ commuting Killing vectors $\xi_i \equiv \partial /
\partial x^i$. For convenience, introduce a new coordinate $\sigma$
by $d\sigma^2=\gamma d\rho^2$ where $\gamma= \det \gamma_{ij}$ and
let $f^2=\gamma\Gamma$. The full spacetime metric $g_{\mu\nu}$ may
be expressed in terms of 3d data: a set of functions $\gamma_{ij}$ ,
a set of one-forms (twist vectors) $\Omega_i = \star (d\xi_i \wedge
\xi_1 \wedge \xi_2 \wedge \cdots \xi_{D-3})$, and an induced metric
$h_{ab}=\gamma( g_{ab}- \gamma^{ij}\xi_{ia}\xi_{jb})$ where the
indices $a,b$ run over all coordinates except $x^i$~\cite{Maison2}. T
The vacuum equations imply that the twist vectors are
closed, $\Omega_i=d\omega_i$ ($\omega_i$ are called the twist
potentials), and leads to a 3d sigma model with equations of motion:
\be R_{ab}(h)= \frac{1}{4}{\rm Tr} (
\partial_a \chi \chi^{-1}
\partial_b\chi \chi^{-1}), \qquad D^a(\partial_a \chi \chi^{-1})=0.
\ee where $D$ is the covariant derivative wrt $h$ and $\chi$ is a
symmetric $(D-2) \times (D-2)$  unimodular
matrix~\cite{Maison1,Maison2}: \be \chi = \left(
\begin{array}{cc} \gamma^{-1} & - \gamma^{-1}\omega_i \\
-\gamma^{-1}\omega_i & \gamma_{ij}+ \gamma^{-1}\omega_i\omega_j
\end{array} \right). \ee In the case we are considering: \be
h_{ab}dx^adx^b= d\sigma^2+ f^2(\sigma)( A_0r^2dv^2+2dvdr), \qquad
\omega_i = \int d\sigma k_i(\sigma)\gamma f^{-2}(\sigma) \ee and the
non-vanishing components of the Ricci tensor of $h$ are: \be
R_{vr}=-\frac{1}{2}\left( \frac{d^2f^2}{d\sigma^2}  - 2A_0 \right),
\qquad R_{\sigma \sigma}= -\frac{1}{2f^4}\left[
2f^2\frac{d^2f^2}{d\sigma^2} - \left(\frac{df^2}{d\sigma}  \right)^2
\right]. \ee Since $\chi$ is independent of $v$, from the sigma
model equations we see that $R_{vr}(h)=0$ and thus we can integrate
to get $f^2$: \bea f^2 = A_0[(\sigma-b)^2-a^2/4] \eea where $a,b$
are constants ($f^2 \ge 0$ implies $a^2 \ge 0$ as $A_0 < 0$).
The field equation
for $\chi$ simplifies as $\chi$ only depends on $\sigma$: \be
\frac{d}{d\sigma} \left( \frac{d\chi}{d\sigma} \chi^{-1} f^2
\right)=0 \ee which can be integrated: \be \chi = \exp[ \mu
R(\sigma) ] \chi_0, \qquad R(\sigma) = \int \frac{d\sigma}{f^2} \ee
where $\mu$ is an arbitrary constant traceless matrix, $\chi_0$ is a
constant unimodular symmetric matrix and $\chi_0 \mu^T =\mu \chi_0$.
The $\sigma\sigma$ component of the Ricci equation then implies
$\textrm{Tr} \mu^2 = 2A_0^2a^2$.
 Performing the
integration explicitly gives: \bea R(\sigma) &=&  \frac{1}{A_0 a}
\log \left| \frac{\sigma-b-a/2}{\sigma-b+a/2} \right|.
\eea Thus we have completely determined the 3d data $h_{ab}$ and
$\chi$, and hence a general vacuum near-horizon geometry, in terms
of the constant matrices $\mu, \chi_0$ and two other integration
constants $a,b$ subject to the constraints derived above. Note that
$\chi \to M \chi M^T$ where $M$ is in $SL(D-2,R)$ leaves the 3d
field equations invariant. On our solution for $\chi$ this freedom
reads $\chi_0 \to M\chi_0 M^T$ and $\mu \to M \mu M^{-1}$. However,
solutions to the 3d equations related by this symmetry will not in
general lead to equivalent spacetime geometries in $D$ dimensions.
However, a subgroup of these transformations which does lead to
equivalent $D$ dimensional geometries is the $GL(D-3,R)$ group which
mixes the the $D-3$ Killing vectors.

The above analysis is local: compactness of the horizon has not been enforced. This will impose further restrictions. In the $D=4$ case, it is known that the general axisymmetric vacuum near-horizon solution with $S^2$ topology is that of the extremal Kerr solution \cite{isolated}.

\section{Discussion}

There are various ways in which our results could be extended.
For example, Theorem 2 assumes the black hole is
a black hole to lowest order (i.e. within Einstein gravity). Thus
our result does not apply to ``small" black holes which are black
holes only when higher derivative terms are taken into account. It
would be nice to extend our proof to remove this assumption. However, this may depend on the details of precisely which higher-derivative terms are required.

In five dimensions, our results assume two rotational symmetries, whereas the ``stationary implies axisymmetric" theorem guarantees only one. It would be interesting to see whether one could extend our results in five dimensions by removing the assumption of this extra rotational symmetry.

We have considered a theory of gravity coupled to abelian vectors and uncharged scalars. Can our results be generalized to theories with non-abelian vectors and/or charged scalars?

We have commented on the special case of a static near-horizon geometry with local $AdS_3$ symmetry. As we have seen, in this case the near-horizon geometry is a warped product of $AdS_3$ and  $S^2$ and the horizon topology $S^1 \times S^2$, i.e., a black ring. Of particular relevance to string theory is the question of whether this structure will be preserved by higher-derivative corrections, or whether it will be broken to the global $AdS_2$ symmetry that we have shown must always exist.\footnote{
See \cite{dab} for a complementary discussion of this point.}
 One can attempt to modify the argument of section \ref{sec:hd} to prove that $AdS_3$ symmetry must exist to all orders in $\lambda$ if it exists for $\lambda=0$ but this does not work.\footnote{To see what goes wrong, consider (for simplicity) a theory of pure gravity. The argument of section \ref{sec:hd} consisted of 2 steps. First, we showed that if the metric is $SO(2,1)$-invariant to order $n$ in $\lambda$ then the RHS of the Einstein equation is $SO(2,1)$-invariant to order $n+1$, hence the Einstein tensor is $SO(2,1)$-invariant to order $n+1$. Second, we showed that this implies that the metric is $SO(2,1)$-invariant to order $n+1$. The second step doesn't work for $O(2,2)$ symmetry, i.e., an $O(2,2)$-symmetric Einstein tensor does not imply an $O(2,2)$-symmetric metric. This is obvious even at zeroth order: the RHS of the Einstein equation is zero, which is obviously $O(2,2)$-symmetric, but this does not imply that any vacuum near-horizon metric must be $O(2,2)$-symmetric, in fact none is! (This follows from \cite{crt}.)} The problem is that $AdS_2$-symmetry is a consequence of the equations of motion, whereas $AdS_3$ appears to be an ``accident" that arises when a non-static black ring solution happens to have a static near-horizon geometry. In general, there is no reason why this accidental feature should persist in the presence of higher-derivative terms.

This conclusion may be modified if one imposes additional symmetries on the solution. For example, in supergravity theories one can impose the condition that the black hole be supersymmetric, and supersymmetry may then explain the ``accidental" $AdS_3$.
For example, in minimal 5d supergravity, the only supersymmetric near-horizon solution with $S^1 \times S^2$ topology is $AdS_3 \times S^2$ \cite{hsr}, corresponding to the near-horizon geometry of a supersymmetric black ring \cite{bpsring}. It would be interesting to extend the classification \cite{hsr} of supersymmetric near-horizon geometries to include higher-derivative terms to see whether this conclusion persists in a more general theory.

\bigskip
\begin{center}{\bf{Acknowledgments}} \end{center}

HKK and JL are supported by PPARC. HSR is a Royal Society University
Research Fellow.

\appendix

\section{Excluding near-horizon geometries with $A_0 \geq 0$}

In this section we wish to show that there are no near-horizon
geometries with compact horizons with $A_0 \geq 0$ in the general
second order theory (\ref{gentheory}). We will assume $V(\phi)\leq
0$, and that $f_{AB},g_{IJ}$ are positive definite.

For reference, the metric is: \be ds^2=\Gamma(\rho)[A_0 r^2 dv^2+2
dvdr ] +d\rho^2 + \gamma_{ij}(\rho)(dx^i+k^irdv)(dx^j+k^jrdv) \ee
and recall that $\Gamma>0$. The Maxwell fields are given by \be
 F^I= \Delta^I (\rho)\, dv \wedge dr + B^I_i(\rho) \,d\rho \wedge (dx^i+k^irdv)
\ee where we have defined the functions $\Delta^I \equiv
F^I_{vr}(\rho)$ and $B^I_i \equiv F^I_{\rho i}(\rho)$ for notational
convenience.

We can now generalise an argument of Gibbons \cite{gibbons}. For the
above metric, we have \be
 R_{vr} = A_0+ \frac{k^ik_i}{2\Gamma}-\frac{1}{2} \left[ \Gamma''+ \frac{\Gamma'
\gamma'}{2\gamma}\right] = A_0+\frac{k^ik_i}{2\Gamma}- \frac{1}{2}
\hat{\nabla}^2 \Gamma, \ee where $\hat{\nabla}$ is the metric
connection on $H$. The $vr$ component of the Einstein equation is
\be A_0 - \frac{1}{2} \hat{\nabla}^2 \Gamma =
-\frac{k^ik_i}{2\Gamma}+\frac{1}{D-2} \Gamma V -
\frac{(D-3)}{2(D-2)} \Gamma^{-1} g_{IJ} \Delta^I \Delta^J -
\frac{1}{2(D-2)} \Gamma g_{IJ} \gamma^{ij}  B_i^I B_j^J. \ee Note
that our assumptions imply that the RHS is non-positive: this is a
consequence of this theory obeying the strong energy condition.
Integrate this equation over $H$ to conclude that $A_0 \leq 0$ which
rules out the case $A_0>0$. Further, if $A_0=0$ we must have $k^i=0$
and $\Delta^I \equiv 0$, $B^I_i \equiv 0$, i.e., the Maxwell fields
vanish, and $V=0$ everywhere. If the theory has $V<0$ then this is a
contradiction and we are done. Otherwise we conclude that the
scalars must take values such that the potential is at its maximum
$V=0$ everywhere. The above equation then tell us that $\Gamma$ is
harmonic, and hence must be constant. Therefore the geometry is a
direct product of $R^{1,1}$ and $H$.

The next step is to show that the scalars must be constant.\footnote{If the assumption that the scalars are invariant under the rotational symmetries were relaxed then this needn't be true \cite{gellmann}.} The scalar equation of motion admits the integral
\be
 \frac{1}{2} f_{AB} {\phi^A}' {\phi^B}' = E,
\ee
where $E$ is a (non-negative) constant and we have used $V=0$. Hence if we can show $E=0$ then the scalars are constant. From the Einstein equation we find
\be
 E = R_{\rho \rho} - \gamma^{ij} R_{ij} = \frac{1}{4} \left[ \frac{{\gamma'}^2}{\gamma^2} - \gamma^{ij} \gamma^{kl} \gamma'_{ik} \gamma'_{jl} \right],
\ee
where $\gamma^{ij}$ is the inverse of $\gamma_{ij}$, $\gamma \equiv \det \gamma_{ij}$, and we have used the fact that $\Gamma$ is constant.  If $D=4$ then the RHS vanishes identically and hence $E=0$. For $D=5$ we can argue as follows. As explained in the main text, $\rho$ takes values on an interval such that $\gamma$ is positive on the interior of the interval and vanishes at the endpoints. Hence there must be a point in the interior of this interval for which $\gamma'=0$. Evaluate the above equation at this point. The RHS is manifestly non-positive, but the LHS is non-negative. Hence we must have $E=0$. Therefore the scalars are constant.

We have shown that the scalars are constant, the Maxwell fields vanish, $V=0$, and $\Gamma$ is constant. Hence the Einstein equation reduces to the vacuum Einstein equation. The metric is a direct product of 2-dimensional flat space with $H$ so the Einstein equation implies that $H$ is Ricci-flat and hence flat (as $H$ is 2 or 3-dimensional). Therefore $H$ must be a torus, contradicting the assumption of Theorem 1.


\begin{thebibliography}{99}


\bibitem{FKS}
  S.~Ferrara, R.~Kallosh and A.~Strominger,
  ``N=2 extremal black holes,''
  Phys.\ Rev.\  D {\bf 52} (1995) 5412
  [arXiv:hep-th/9508072].

\bibitem{Strominger}
  A.~Strominger,
  ``Macroscopic Entropy of $N=2$ Extremal Black Holes,''
  Phys.\ Lett.\  B {\bf 383} (1996) 39
  [arXiv:hep-th/9602111].

\bibitem{FK}
  S.~Ferrara and R.~Kallosh,
  ``Supersymmetry and Attractors,''
  Phys.\ Rev.\  D {\bf 54} (1996) 1514
  [arXiv:hep-th/9602136].

\bibitem{hd1}
 G.~Lopes Cardoso, B.~de Wit and T.~Mohaupt,
  ``Corrections to macroscopic supersymmetric black-hole entropy,''
  Phys.\ Lett.\  B {\bf 451} (1999) 309
  [arXiv:hep-th/9812082].

\bibitem{hd2}
  G.~Lopes Cardoso, B.~de Wit and T.~Mohaupt,
  ``Macroscopic entropy formulae and non-holomorphic corrections for
  supersymmetric black holes,''
  Nucl.\ Phys.\  B {\bf 567} (2000) 87
  [arXiv:hep-th/9906094].

\bibitem{hd3}
 G.~Lopes Cardoso, B.~de Wit, J.~Kappeli and T.~Mohaupt,
  ``Stationary BPS solutions in N = 2 supergravity with R**2 interactions,''
  JHEP {\bf 0012} (2000) 019
  [arXiv:hep-th/0009234].

\bibitem{sen1}
  A.~Sen,
  ``Black hole entropy function and the attractor mechanism in higher
  derivative gravity,''
  JHEP {\bf 0509}, 038 (2005)
  [arXiv:hep-th/0506177].

\bibitem{nonsusyatt}
  K.~Goldstein, N.~Iizuka, R.~P.~Jena and S.~P.~Trivedi,
  ``Non-supersymmetric attractors,''
  Phys.\ Rev.\  D {\bf 72} (2005) 124021
  [arXiv:hep-th/0507096].

\bibitem{Kallosh}
  R.~Kallosh,
  ``New attractors,''
  JHEP {\bf 0512} (2005) 022
  [arXiv:hep-th/0510024].

\bibitem{agm}
  D.~Astefanesei, K.~Goldstein and S.~Mahapatra,
  ``Moduli and (un)attractor black hole thermodynamics,''
  arXiv:hep-th/0611140.

\bibitem{dab}
A.~Dabholkar, A.~Sen and S.~P.~Trivedi,
  ``Black hole microstates and attractor without supersymmetry,''
  JHEP {\bf 0701}, 096 (2007)
  [arXiv:hep-th/0611143].

\bibitem{KLMS}
  D.~M.~Kaplan, D.~A.~Lowe, J.~M.~Maldacena and A.~Strominger,
  ``Microscopic entropy of N = 2 extremal black holes,''
  Phys.\ Rev.\  D {\bf 55} (1997) 4898
  [arXiv:hep-th/9609204]

\bibitem{HLM}
  G.~T.~Horowitz, D.~A.~Lowe and J.~M.~Maldacena,
  ``Statistical Entropy of Nonextremal Four-Dimensional Black Holes and
  U-Duality,''
  Phys.\ Rev.\ Lett.\  {\bf 77} (1996) 430
  [arXiv:hep-th/9603195].

\bibitem{Dabholkar}
  A.~Dabholkar,
  ``Microstates of non-supersymmetric black holes,''
  Phys.\ Lett.\  B {\bf 402} (1997) 53
  [arXiv:hep-th/9702050].

\bibitem{TT}
  P.~K.~Tripathy and S.~P.~Trivedi,
  ``Non-supersymmetric attractors in string theory,''
  JHEP {\bf 0603} (2006) 022
  [arXiv:hep-th/0511117].

\bibitem{EH}
  R.~Emparan and G.~T.~Horowitz,
  ``Microstates of a neutral black hole in M theory,''
  Phys.\ Rev.\ Lett.\  {\bf 97} (2006) 141601
  [arXiv:hep-th/0607023].

\bibitem{EM}
R.~Emparan and A.~Maccarrone,
  ``Statistical description of rotating Kaluza-Klein black holes,''
  Phys.\ Rev.\  D {\bf 75}, 084006 (2007)
  [arXiv:hep-th/0701150].



\bibitem{rotatt}
D.~Astefanesei, K.~Goldstein, R.~P.~Jena, A.~Sen and S.~P.~Trivedi,
  ``Rotating attractors,''
  JHEP {\bf 0610}, 058 (2006)
  [arXiv:hep-th/0606244].


\bibitem{hsr}
  H.~S.~Reall,
  ``Higher dimensional black holes and supersymmetry,''
  Phys.\ Rev.\  D {\bf 68} (2003) 024024
  [Erratum-ibid.\  D {\bf 70} (2004) 089902]
  [arXiv:hep-th/0211290].


\bibitem{BH}
  J.~M.~Bardeen and G.~T.~Horowitz,
  ``The extreme Kerr throat geometry: A vacuum analog of AdS(2) x S(2),''
  Phys.\ Rev.\ D {\bf 60} (1999) 104030 [arXiv:hep-th/9905099].

\bibitem{MyersPerry}
  R.~C.~Myers and M.~J.~Perry,
  ``Black Holes In Higher Dimensional Space-Times,''
  Annals Phys.\  {\bf 172} (1986) 304.



\bibitem{ring}
R.~Emparan and H.~S.~Reall,
  Phys.\ Rev.\ Lett.\  {\bf 88}, 101101 (2002)
  [arXiv:hep-th/0110260].


\bibitem{Pomeransky}
  A.~A.~Pomeransky and R.~A.~Sen'kov,
  ``Black ring with two angular momenta,''
  arXiv:hep-th/0612005.

\bibitem{Emparan}
  R.~Emparan,
  ``Rotating circular strings, and infinite non-uniqueness of black rings,''
  JHEP {\bf 0403} (2004) 064
  [arXiv:hep-th/0402149].

\bibitem{hawkingellis}
S.W. Hawking and G.F.R. Ellis, ``The large scale structure of space-time", Cambridge University Press (1973).

\bibitem{Wald}
  S.~Hollands, A.~Ishibashi and R.~M.~Wald,
  ``A higher dimensional stationary rotating black hole must be
  axisymmetric,''
  Commun.\ Math.\ Phys.\  {\bf 271}, 699 (2007)
  [arXiv:gr-qc/0605106]

\bibitem{chruscielwald}
 P.~T.~Chrusciel and R.~M.~Wald,
  ``On The Topology Of Stationary Black Holes,''
  Class.\ Quant.\ Grav.\  {\bf 11}, L147 (1994)
  [arXiv:gr-qc/9410004].

\bibitem{gallowaycensor}
G.~J.~Galloway, K.~Schleich, D.~M.~Witt and E.~Woolgar,
  ``Topological censorship and higher genus black holes,''
  Phys.\ Rev.\  D {\bf 60}, 104039 (1999)
  [arXiv:gr-qc/9902061].

\bibitem{Galloway}
  G.~J.~Galloway and R.~Schoen,
  ``A generalization of Hawking's black hole topology theorem to higher
  dimensions,''
  Commun.\ Math.\ Phys.\  {\bf 266} (2006) 571
  [arXiv:gr-qc/0509107].
G.~J.~Galloway,
  ``Rigidity of outer horizons and the topology of black holes,''
  arXiv:gr-qc/0608118.

\bibitem{SBH1}
  A.~Dabholkar,
  ``Exact counting of black hole microstates,''
  Phys.\ Rev.\ Lett.\  {\bf 94} (2005) 241301
  [arXiv:hep-th/0409148].

\bibitem{SBH2}
  A.~Dabholkar, R.~Kallosh and A.~Maloney,
  ``A stringy cloak for a classical singularity,''
  JHEP {\bf 0412} (2004) 059
  [arXiv:hep-th/0410076].

\bibitem{SBH3}
  V.~Hubeny, A.~Maloney and M.~Rangamani,
  ``String-corrected black holes,''
  JHEP {\bf 0505} (2005) 035
  [arXiv:hep-th/0411272].

\bibitem{SBH4}
  A.~Dabholkar, F.~Denef, G.~W.~Moore and B.~Pioline,
  ``Precision counting of small black holes,''
  JHEP {\bf 0510} (2005) 096
  [arXiv:hep-th/0507014].

\bibitem{SBH5}
  A.~Sen,
  ``Stretching the horizon of a higher dimensional small black hole,''
  JHEP {\bf 0507} (2005) 073
  [arXiv:hep-th/0505122].

\bibitem{Maison2}
  P.~Dobiasch and D.~Maison,
  ``Stationary, Spherically Symmetric Solutions Of Jordan's Unified Theory Of
  Gravity And Electromagnetism,''
  Gen.\ Rel.\ Grav.\  {\bf 14} (1982) 231.

\bibitem{racz}
H.~Friedrich, I.~Racz and R.~M.~Wald,
  ``On the Rigidity Theorem for Spacetimes with a Stationary Event Horizon or a
  Compact Cauchy Horizon,''
  Commun.\ Math.\ Phys.\  {\bf 204}, 691 (1999)
  [arXiv:gr-qc/9811021].

\bibitem{bpsring}
  H.~Elvang, R.~Emparan, D.~Mateos and H.~S.~Reall,
  ``A supersymmetric black ring,''
  Phys.\ Rev.\ Lett.\  {\bf 93}, 211302 (2004)
  [arXiv:hep-th/0407065].

\bibitem{gowdy}
  R.~H.~Gowdy,
  ``Vacuum space-times with two parameter spacelike isometry groups and compact
  invariant hypersurfaces: Topologies and boundary conditions,''
  Annals Phys.\  {\bf 83} (1974) 203.


\bibitem{KLR}
  H.~K.~Kunduri, J.~Lucietti and H.~S.~Reall,
  ``Do supersymmetric anti-de Sitter black rings exist?,''
  JHEP {\bf 0702} (2007) 026
  [arXiv:hep-th/0611351].

\bibitem{crt}
  P.~T.~Chrusciel, H.~S.~Reall and P.~Tod,
  ``On non-existence of static vacuum black holes with degenerate  components
  of the event horizon,''
  Class.\ Quant.\ Grav.\  {\bf 23} (2006) 549
  [arXiv:gr-qc/0512041].

\bibitem{Maison1}
  D.~Maison,
  ``Ehlers-Harrison Type Transformations For Jordan's Extended Theory Of
  Gravitation,''
  Gen.\ Rel.\ Grav.\  {\bf 10} (1979) 717.

\bibitem{HHT}
  S.~W.~Hawking, C.~J.~Hunter and M.~M.~Taylor-Robinson,
  ``Rotation and the AdS/CFT correspondence,''
  Phys.\ Rev.\  D {\bf 59} (1999) 064005
  [arXiv:hep-th/9811056].

\bibitem{Chong}
  Z.~W.~Chong, M.~Cvetic, H.~Lu and C.~N.~Pope,
  ``General non-extremal rotating black holes in minimal five-dimensional
  gauged supergravity,''
  Phys.\ Rev.\ Lett.\  {\bf 95}, 161301 (2005)
  [arXiv:hep-th/0506029].

\bibitem{KLR2}
  H.~K.~Kunduri, J.~Lucietti and H.~S.~Reall,
  ``Supersymmetric multi-charge AdS(5) black holes,''
  JHEP {\bf 0604} (2006) 036
  [arXiv:hep-th/0601156].


\bibitem{SabraKlemm}
  D.~Klemm and W.~A.~Sabra,
  ``Charged rotating black holes in 5d Einstein-Maxwell-(A)dS gravity,''
  Phys.\ Lett.\  B {\bf 503}, 147 (2001)
  [arXiv:hep-th/0010200].


\bibitem{ringreview}
  R.~Emparan and H.~S.~Reall,
  Class.\ Quant.\ Grav.\  {\bf 23} (2006) R169
  [arXiv:hep-th/0608012].

\bibitem{LPV}
  H.~Lu, C.~N.~Pope and J.~F.~Vazquez-Poritz,
  ``From AdS black holes to supersymmetric flux-branes,''
  Nucl.\ Phys.\  B {\bf 709} (2005) 47
  [arXiv:hep-th/0307001].


\bibitem{Gauntlett}
  J.~P.~Gauntlett, N.~Kim and D.~Waldram,
  ``Supersymmetric AdS(3), AdS(2) and bubble solutions,''
  arXiv:hep-th/0612253.

\bibitem{IM}
  H.~Ishihara and K.~Matsuno,
  Prog.\ Theor.\ Phys.\  {\bf 116} (2006) 417
  [arXiv:hep-th/0510094].


\bibitem{isolated}
J.~Lewandowski and T.~Pawlowski,
  Class.\ Quant.\ Grav.\  {\bf 20}, 587 (2003)
  [arXiv:gr-qc/0208032].

\bibitem{gibbons}
G.~W.~Gibbons,
 ``Aspects Of Supergravity Theories,''  lectures given at GIFT Seminar on Theoretical Physics, San Feliu de Guixols, Spain, Jun 4-11, 1984.

\bibitem{gellmann}
M.~Gell-Mann and B.~Zwiebach,
  Phys.\ Lett.\  B {\bf 141}, 333 (1984).



\end{thebibliography}
\end{document}